\documentclass[twocolumn,showkeys,showpacs,preprintnumbers,prd,superscriptaddress,nofootinbib]{revtex4-1}
\usepackage{graphicx}	
\usepackage{amssymb}
\usepackage{dcolumn}
\usepackage{mathtools}
\usepackage{amsmath}
\usepackage{xcolor}
\usepackage{color}
\usepackage{subfigure, rotating, bm, array}
\usepackage[pagebackref=false, colorlinks=true]{hyperref}
\hypersetup{linkcolor=blue, citecolor=blue,urlcolor=blue}

\begin{document}

\title{High Energy Particle Collisions in the vicinity of Naked Singularity}


\author{Kauntey Acharya}
\email{kaunteyacharya2000@gmail.com}
\affiliation{International Centre for Space and Cosmology, Ahmedabad University, Ahmedabad, GUJ 380009, India}
\author{Parth Bambhaniya}
\email{grcollapse@gmail.com}
\affiliation{Instituto de Astronomia, Geofísica e Ciências Atmosféricas, Universidade de São Paulo, IAG, Rua do Matão 1225, CEP: 05508-090 São Paulo - SP - Brazil.}
\author{Pankaj S. Joshi}
\email{psjcosmos@gmail.com}
\affiliation{International Centre for Space and Cosmology, Ahmedabad University, Ahmedabad, GUJ 380009, India}
\author{Kshitij Pandey}
\email{pkshitij45@gmail.com}
\affiliation{PDPIAS, Charusat University, Anand-388421 (Gujarat), India}
\affiliation{International Center for Cosmology, Charusat University, Anand, Gujarat 388421, India}
\author{Vishva Patel}
\email{vishwapatel2550@gmail.com}
\affiliation{PDPIAS, Charusat University, Anand-388421 (Gujarat), India}
\affiliation{International Center for Cosmology, Charusat University, Anand, Gujarat 388421, India}

\date{\today}

\begin{abstract}
In this paper, we investigate particle acceleration and high-energy collisions in the Joshi-Malafarina-Narayan (JMN-1) naked singularity, which, in the absence of an event horizon, allows infalling particles to turn back under specific angular momentum conditions. These outgoing particles can then collide with infalling ones, enabling the JMN-1 singularity to act as a natural high-energy particle accelerator. We derive the necessary expressions to compute the center-of-mass energy of two colliding particles and find that this energy can reach extremely high values, potentially even approaching Planck energy scales. We also explore the implications of these results, including the possible formation of microscopic black holes that could decay via Hawking radiation, releasing energy on the order of  $10^{26} eV$ due to the extreme gravitational fields near the naked singularity. This scenario offers significant advantages. If horizonless compact objects exist in nature, these high-energy collisions could substantially influence the surrounding physical processes and might give rise to distinct observational signatures.

\end{abstract}
\maketitle

\section{Introduction}
\label{sec_intro}
The ultra-compact objects such as black holes, naked singularities, gravastars and others are under much investigation today due to observed extreme high energy phenomena in the universe such as the active galactic nuclei, black hole mergers, gamma-ray bursts and such others. On theoretical side, the gravitational collapse of massive matter clouds that would typically give rise to such ultra-compact objects has been examined extensively. It is known now that the visible and hidden singularities do form from the continual gravitational collapse from regular initial data in general relativity, for example for massive stars collapsing catastrophically at the end of their life-cycles \cite{joshi,goswami,mosani1,mosani2,mosani3,mosani4,Deshingkar:1998ge,Jhingan:2014gpa,Joshi:2011zm}. 
In particular, it is also found that the naked singularities of collapse do develop generically under a wide variety of physically reasonable conditions \cite{Broderick:2024vjp,Joshi:2024gog,Joshi:2011rlc}.

The recent findings of the EHT collaboration made significant progress in studying the black hole event horizon, advancing our understanding and suggesting that the evidence does not rule out Sgr A* being a JMN-1 naked singularity \cite{EventHorizonTelescope:2022xqj}. The accretion disks and shadow properties of naked singularities have been studied \cite{Joshi:2013dva,Saurabh:2023otl,gyulchev,Bambhaniya:2021ugr,Tahelyani:2022uxw,Kovacs:2010xm,Guo:2020tgv,Chowdhury:2011aa,Joshi:2020tlq,Saurabh:2022jjv,Solanki:2021mkt,shaikh1,Bambhaniya:2021ybs,Bambhaniya:2024lsc,Vagnozzi:2022moj,Patel:2022vlu}. Other observational signatures including gravitational lensing \cite{Virbhadra:2007kw,Gyulchev:2008ff,Sahu:2012er}, energy extractions \cite{Patel:2023efv,Patel:2022jbk,Viththani:2024map}, relativistic orbits \cite{Martinez,Bambhaniya:2019pbr,Joshi:2019rdo,tsirulev,Dey:2019fpv,Bam2020,Bambhaniya:2025xmu,Bambhaniya:2022xbz,Dey:2020haf,Bambhaniya:2021jum}, tidal forces \cite{Madan:2022spd,Viththani:2024fod,Arora:2023ltv,Joshi:2024djy}, and time delay of pulsar signals \cite{Kalsariya:2024qyp} are explored in the naked singularity spacetimes.

The collision of ultra-high energy particles near such ultra-compact objects through particle acceleration in extreme strong gravitational fields present in their vicinity can be one of the most prominent sources of observed high energy phenomena in the universe. Therefore, in this paper, we aim to explore the mechanisms behind particle acceleration in the vicinity of ultra-compact objects, in particular for the JMN-1 naked singularity spacetime. Specifically, we focus here on timelike geodesics in the JMN-1 geometry to understand the particle acceleration process. Our approach does not consider the role of radiation or fields in the acceleration process, allowing us to isolate and analyze the effect of geodesic motions on particle acceleration. Several earlier studies have shown the particle acceleration phenomena around ultra-compact objects. In \cite{Patil:2011yb}, authors showed that for the galactic object Sgr A*, the collisional energy of two particles can reach $10^{3}\,TeV$. 

The phenomenon of particle acceleration and collisions around black holes is known as the Banados-Silk-West effect \cite{Banados:2009pr}.
It is to be noted, however, that this effect required finely-tuned scenarios for the collision of particles to release energy, which makes it less appealing as a mechanism for explaining high-energy astrophysical phenomena. Following this, authors in \cite{Patil:2011aa,Patil:2011uf,Patil:2011ya} studied the particle acceleration process in the vicinity of the naked singularity which requires less fine tuning for the possibility of high energy collisions of particles. The intense gravitational region of ultracompact object can accelerate particles to energies near the Planck scale, as shown in \cite{Patil:2011aw}. Therefore, in the present work, we consider particle acceleration as a potential mechanism to explain the high energy phenomena by considering the collision of particles in the vicinity of JMN-1 naked singularity. We investigate the behavior of energy of colliding particles in the center of mass frame $(E_{CM})$ with respect to radial distance as well as for the angular momenta of two particles. Further, we also derive the numerical value of the center of mass energy $(E_{CM})$ by considering JMN-1 naked singularity at the milky-way galactic center.

This paper is organized as follows: The JMN-1 naked singularity is discussed in section (\ref{sec_jmn1}). The geodesic motion of a particle and the center of mass energy around the JMN-1 naked singularity are derived in section (\ref{sec_particle}). In the section (\ref{sec_microbh}), we discuss the possibility of formation of microscopic black holes. This possibility is described with a discussion of Planck scale energy and Hawking radiation for the high center of mass energy coming from the particles collision. Finally, we conclude and discuss the results of this work in section (\ref{sec_conclusion}). Only when necessary, we defined the gravitational constant (G) and speed of light (c) values, otherwise these are defined as geometrized units with unity values. 

\section{Joshi–Malafarina–Narayan (JMN-1) spacetime}
\label{sec_jmn1}
The very first model of gravitational collapse of a spherically symmetric and homogeneous dust cloud was proposed by Oppenheimer, Snyder, and Dutt, (OSD) which indeed indicates the formation of a black hole as an end state of the collapse \cite{Oppenheimer:1939ue}. In this OSD model of gravitational collapse, trapped surfaces form around the center before the formation of the central space-like singularity. As a result, the central singularity is causally disconnected from other points of spacetime, indicating the presence of an event horizon in the spacetime structure. However, the OSD model assumes a completely homogeneous density and zero pressure within the massive collapsing star together with other simplifying assumptions. Thus it is considered to be a rather idealistic scenario of gravitational collapse.

Based on the above OSD model, Roger Penrose proposed the Cosmic Censorship Conjecture (CCC), which hypothesized that gravitational collapse must end in a black hole formation only, where the final spacetime singularity is necessarily hidden within an event horizon of gravity
\cite{penrose}.
However, despite numerous attempts to formulate the same mathematically or to prove the CCC, there has been no success in settling this one of the most important issues in gravitation physics, which is at the very foundation of all of black hole physics and its astrophysical applications.
Therefore several investigations have been carried out to understand the process of gravitational collapse in more realistic physical scenarios in Einstein gravity, including the inhomogeneous distribution of matter in the collapsing cloud and non-zero pressure profiles \cite{joshi,goswami,mosani1,mosani2,mosani3,mosani4,Deshingkar:1998ge,Jhingan:2014gpa,Joshi:2011zm,Joshi:2011rlc}.

In \cite{Joshi:2011zm}, authors have extensively studied equilibrium configurations of the collapsing cloud under the influence of gravity with zero radial pressure but with non-zero tangential pressures, i.e. an anisotropic fluid with pressure. It was shown that non-zero tangential pressure can prevent the formation of trapped surfaces around the central region. An end state of a massive star under such conditions can lead to the formation of JMN-1 naked singularity. The line element of the JMN-1 naked singularity spacetime is,
\begin{equation}
     ds^2 = - (1- M_{0}) \left( \frac{r}{R_b}\right)^{\frac{M_{0}}{1 - M_{0}}} dt^2 + \frac{1}{(1- M_{0})}  dr^2 +\, r^2  d\Omega^2,
     \label{JNWst}
\end{equation}
where, $d\Omega^2= d\theta^2+\,sin^2\,\theta\,d\phi^2$. $M_{0}$ and $R_{b}$ are positive constants. Here, $M_{0}$ can have any value within the range $0<M_{0}\,<\,4/5$. An upper limit of $M_0$ is given by the fact that the sound speed cannot exceed unity \cite{Joshi:2011zm}. $R_{b}$ is the radius of equilibrium configuration of matter around the central singularity. Note that, $R_b$ should not be less than $2.5 M$ to satisfy the sound speed condition.

The spacetime metric is modeled by considering a high-density compact region in a vacuum, which means that the spacetime configuration should be asymptotically flat. For this purpose, the JMN-1 spacetime is matched with an exterior Schwarzschild spacetime
at radius $r=R_{b}$.
Now for matching of these two spacetimes, there are two junction conditions which suggest that (i) extrinsic curvatures of both spacetimes should smoothly match at a null hypersurface, and (ii) induced metrics of exterior and interior geometries be equivalent on the hypersurface where matching is considered. Since the radial pressure is zero in JMN-1 naked singularity spacetime, the extrinsic curvatures of the interior JMN-1 spacetime and the exterior Schwarzschild spacetime are smoothly matched at $r=R_{b}$ \cite{Bambhaniya:2019pbr}. 
The line element of this exterior spacetime can be written as,
\begin{equation}
    ds^2=-\left(1-\frac{M_{0}R_{b}}{r}\right)dt^2+\frac{dr^2}{\left(1-\frac{M_{0}R_{b}}{r}\right)}+r^2d\Omega^2,
    \label{extsch}
\end{equation}
where, the matching condition is $M_{0}R_{b}=2M$.
It is also important to note that all the energy conditions are satisfied in the above spacetime structure. In the next section, we analyze high-energy particle collisions near the JMN-1 naked singularity.

\section{High energy particle collisions near the JMN-1 naked singularity}
\label{sec_particle}
In this section, we describe the mechanism of particle acceleration and investigate the high energy collisions of particles near the central singularity of JMN-1 spacetime. We first consider the geodesic motion of colliding particles with turning points and derive the expression of center of mass energy. For this, we also study the allowed values of metric parameter $M_{0}$ for which  energy collisions are possible because of the existence of turning points. 

\subsection{Geodesic motion and particle acceleration in the JMN-1 spacetime}
\label{sec_geodesiceqn}
We first describe the motion of the massive particles of mass $m$ moving with the four velocities $U^{\mu}$ in the JMN-1 spacetime. For simplicity, we have considered an equatorial plane ($\theta\,=\,\pi/2$) for particles motion, which suggests that $U^{\theta}=0$. The equations of motion of particles can be written by using constants of motion, i.e. conserved energy and conserved angular momentum per unit rest mass of the particle. From these constants of motion, the expressions of four velocities are given as,
\begin{eqnarray}
    && \left(\frac{dt}{d\tau}\right)\,=\,- \frac{e}{(1- M_{0})} \left( \frac{r}{Rb}\right)^{- \frac{M_{0}}{1 - M_{0}}},\\
    && \left(\frac{d\phi}{d\tau}\right)=\frac{L}{r^{2}}.
\end{eqnarray}
Using the normalization condition of four velocity for a test particle $u^{\mu}u_{\mu}=-1$, the total relativistic energy expression can be written as,
\begin{equation}
    {\left(\frac{dr}{d\tau}\right)^2}\,+\,V_{eff}(r)\,=\,E,
\end{equation}
where, $E=\frac{e^2-1}{2}$ is the total energy and $V_{eff}(r)$ is the effective potential, which can be expressed as,
 \begin{equation}
 V_{eff} (r) =  \frac{1}{2}\bigg[(1\;-\;M_{o}) \bigg(\frac{r}{R_{b}}\bigg)^{\frac{\;M_{o}}{1\;-\;M_{o}}}\bigg(1 + \frac{L^2}{r^2}\bigg) - 1\bigg]. 
 \end{equation}

We now consider a scenario where non-relativistic particles at infinity fall inwards under the influence of gravity near the JMN-1 naked singularity. However, the exterior region of the JMN-1 naked singularity is represented by the Schwarzschild metric for $(r>R_b)$. 
Thus it is important to derived equations of motion in the exterior Schwarzschild metric as it is considered that colliding particles are coming from infinity. The expressions of four velocity of a test particle using constants of motion can be written as,
\begin{eqnarray}
     && \left(\frac{dt}{d\tau}\right)\,=\,-\frac{e}{\left(1-\frac{M_{0}R_{b}}{r}\right)},\\
     &&  \left(\frac{d\phi}{d\tau}\right)=\frac{L}{r^{2}}.
\end{eqnarray}
Using the normalization condition $u^{\mu}u_{\mu}=-1$ for timelike geodesic, the radial component of the four velocity in the exterior Schwarzschild geometry can be written as,
\begin{equation}
    \left(\frac{dr}{d\tau}\right)^{2}\,=\,\frac{e^{2}-1}{2} - \frac{1}{2}\bigg[\left(1-\frac{M_{0}R_{b}}{r}\right)\bigg(1 + \frac{L^2}{r^2}\bigg) - 1\bigg],
\end{equation}
and the expression of an effective potential becomes accordingly,
\begin{equation}
     V_{eff} (r) =  \frac{1}{2}\bigg[\left(1-\frac{M_{0}R_{b}}{r}\right)\bigg(1 + \frac{L^2}{r^2}\bigg) - 1\bigg].
\end{equation}
\begin{figure}[ht]
     \includegraphics[width=\columnwidth]{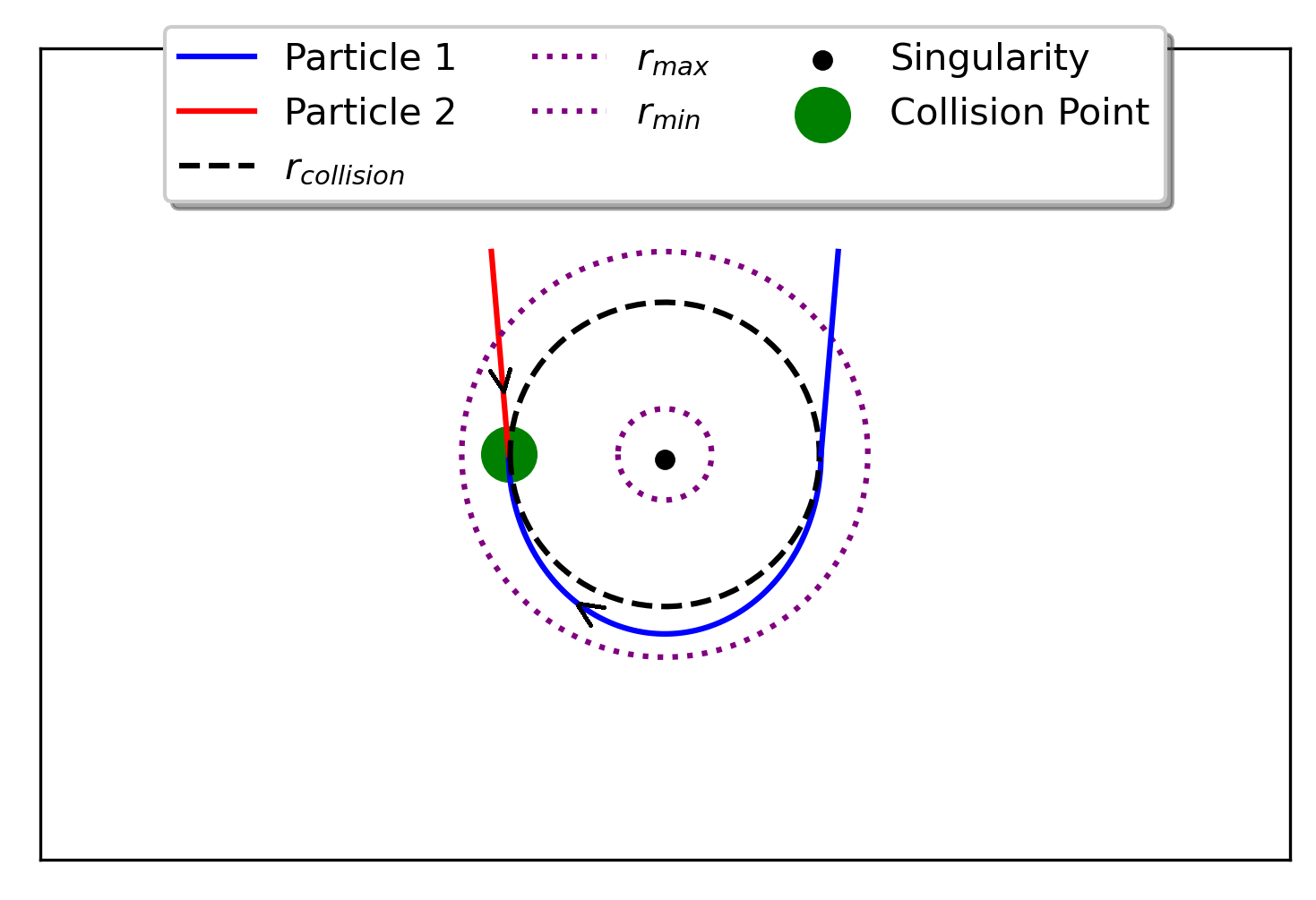}
    \caption {The figure shows that particle 1 is turned at a turning point and collides head-on with particle 2 at a radial distance of $r=3.2M$ from the singularity, corresponding to $M_{0}=0.55$ case. Here $r_{max}$ and $r_{min}$ are the maximum and minimum possible radii for the turning point.}
    \label{fig:my_label}
\end{figure}

For particles at infinity, as $r \rightarrow \infty$, $ U^{r} \rightarrow 0$, the conserved energy per unit rest mass of the particles should be $e=1$.  In general, for high energy collisions, the process must happen near the central singularity, so that very high gravitational effects can be considered. For this to happen, one of the particles must turn back before it reaches near the singularity. For a particle to turn back at any given radial distance $r$, $U^{r}=0$. From this condition one can write the expression of required angular momentum $L$ for the particle to turn back at a radius $r$ in the exterior Schwarzschild geometry,
\begin{equation}
    L=\pm\,r * \left(\frac{ M_{0}R_{b}}{r-M_{0}R_{b}}\right)^\frac{1}{2}.
\end{equation}

However, in order to consider the maximum energy extraction from the collision of particles, one needs to consider this phenomenon of collisions near the singularity. Since the central high-density region is represented by JMN-1 spacetime metric, we consider the collision of particles within this spacetime. Here again, it is considered that the particles must turn back before they could reach the singularity, as shown in the Fig.\,\ref{fig:my_label}. In this geometry, for a particle to turn back at any given radial distance, $U^{r}=0$ and $e=1$ or $E=0$. The expression of required angular momentum $L$ for the particle to turn back at any radial distance $r$ in the JMN-1 spacetime can be written as,
\begin{figure}[ht]
\centering
{\includegraphics[width=\columnwidth]{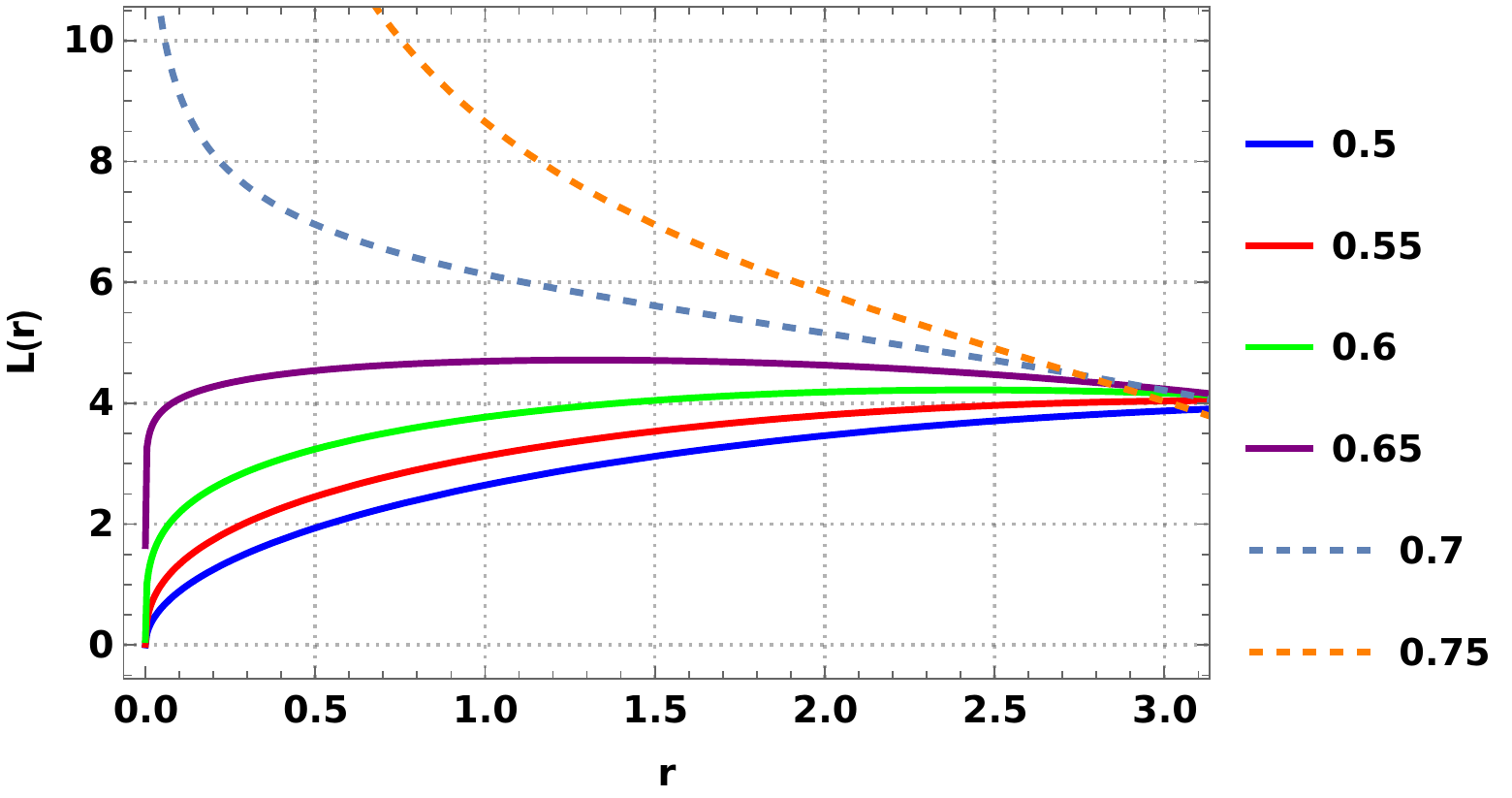}}
\hspace{0.8cm}
 \caption{The figure shows the change in angular momentum ($L(r)$) of a particle with respect to the radial distance ($r$). The color bar on right side of the figure represents the different values of parameter $M_{0}$.}
 \label{fig:jmn1angular}
\end{figure}

\begin{equation}
    L=\pm\,r * \left[\frac{R_{b}^\frac{M_{0}}{1-M_{0}}}{\left(1-M_{0}\right)r^\frac{M_{0}}{1-M_{0}}}-1\right]^{\frac{1}{2}}.\label{angmomjmn}
\end{equation}
The plot of the above expression is shown in Fig.\,\ref{fig:jmn1angular}. It represents the variation in angular momentum for particle to turn back at any radial distance with respect to radius in JMN-1 geometry for different values of $M_{0}$ parameter between $1/2\,<\,M_{0}\,<\,3/4$. Other values of $M_{0}$ are not considered here because there do not exist any turning points for these corresponding values. It should also be noted that for $M_{0}\,<\,1/2$, the radius at which the collision would take place is greater than the equilibrium configuration radius $R_{b}$ of JMN-1. Since the minima in the graph of $L\rightarrow r$ is outside the boundary radius for JMN-1 metric, this range ($M_0<1/2$) is not applicable for particle collisions in the JMN-1.

At the same time, for $M_{0}\,>\,2/3$, the radius at which we get the minimum angular momentum is necessarily imaginary. This can also be seen from the corresponding lines representing the angular momentum of a particle with respect to the distance $r$, that there are no minima in the angular momentum plot for $M_{0}=0.7, 0.75$, This suggests that the particle does not turn back at any radial distance, preventing a collision from occurring. Thus the particle collisions need to be considered only between the parameter range $1/2\,\leq\,M_{0}\,\leq\,2/3$. Corresponding to these values of $M_{0}$, the radial distance at which the collisions would occur can be found using Eq.(\ref{angmomjmn}), which is $r=4.00M$ for $M_{0}=0.50$, $r=3.23M$ for $M_{0}=0.55$, $r=2.43M$ for $M_{0}=0.60$, and $r= 1.31M$ for $M_{0}=0.65$. Here, $M$ is the total mass of the compact object. Therefore, for the case of $M_{0}<\frac{1}{2}$, the radius at which turning point exist outside the boundary radius $R_{b}$, and for $M_{0}>\frac{2}{3}$ the particle will plunge inside the singularity.

Now the range of possible values of angular momentum for the collision to occur is known. Using this information, we can find out the energy of particles after the collisions. However, since the collision is taking place in curved spacetime, there is no fixed or specific coordinate system or reference frame for which the energy of the particles can be defined. Thus we consider a center of mass frame of two particles as a reference frame for this purpose.

\subsection{Center of mass energy in JMN-1 naked singularity}
\label{sec_ecm}
To define the center of mass energy of colliding particles, we first consider a general curved spacetime defined by $g_{\mu \nu}$. In this spacetime, we can consider any arbitrary point as a lab frame where the spacetime can be considered to be locally flat and has a form of Minkowski metric. The local lab frame can be described by basis vectors $\hat{e_{a}}$ for which $\hat{e_{a}} \hat{e_{b}}=\eta_{ab}$. Here for any given basis, the coordinate basis can be written as $\Vec{e_{\mu}}=e^{a}_{\mu} \hat{e_{a}}$. Now if the curved spacetime metric $g_{\mu \nu}$ is given, then the matrix $e^{a}_{\mu}$ can be uniquely defined. Thus from the information about the world-line history $x^{\mu}(\tau)$ and four velocity $u^{\mu}$, three velocity of any particle in that local lab frame can be written as,

\begin{equation}
    v^{(i)}=\frac{e^{(i)}_{\mu} u^{\mu}}{e^{(0)}_{\mu} u^{\mu}}.
\end{equation}
The energy of center of mass in this frame can be expressed as,

\begin{equation}
    E_{cm}^{2}=2m^{2}\left(1-\eta_{\mu \nu}u^{\mu}_{(1)}u^{\nu}_{(2)}\right)^{\frac{1}{2}},
\end{equation}
\begin{figure}[ht]
\centering
\subfigure[$M_{0}$ = 0.55, $R_{b}$ = 3.64, $L_{1} $= 4.05, $L_{2} $= -4.05.]
{\includegraphics[width=\columnwidth]{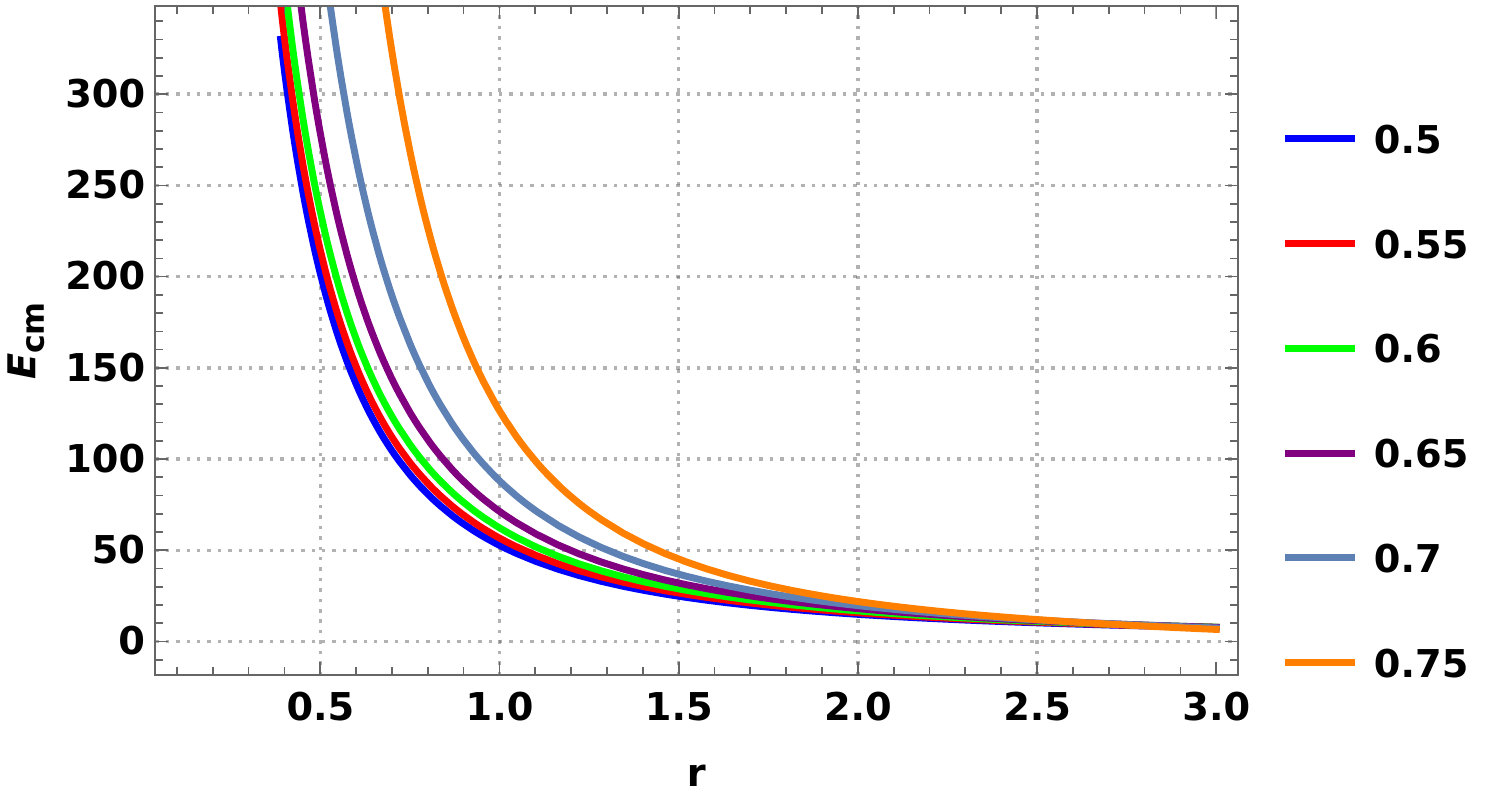}}
\hspace{0.8cm}
\caption{The figure shows change in $E_{cm}$ with radial distance (r). The color bar on right side of the figure represents the different values of parameter $M_{0}$. Here, $L_{1} $= 7, $L_{2} $= -7.}
\label{fig:radiusandecmjmn1}
\end{figure}
the above equation can be written for a generally curved spacetime metric from $g_{\mu \nu}=e^{a}_{\mu} e^{b}_{\nu} \eta_{ab}$ as,
\begin{equation}
     E_{cm}^{2}=2m^{2}\left(1-g_{\mu \nu}u^{\mu}_{(1)}u^{\nu}_{(2)}\right)^{\frac{1}{2}}.
\end{equation}
Now we can use the above equation for exterior Schwarzschild metric and get the following expression of center of mass energy,
\begin{widetext}
\begin{align}
 \left(\frac{ E_{cm}^2}{2m^{2}}\right)_{SCH} &=  1+ \frac{1}{\left(1- \frac{M_{0}R_{b}}{r}\right)}- 
   \frac{1}{\left(1- \frac{M_{0}R_{b}}{r}\right)} \times \nonumber \\ &\left(1-\left(1-\frac{M_{0}R_{b}}{r}\right) \left(1+\frac{L_{1}^{2}}{r^{2}}\right)\right)^{\frac{1}{2}} \left(1-\left(1-\frac{M_{0}R_{b}}{r}\right)\left(1+\frac{L_{2}^{2}}{r^{2}}\right)\right)^{\frac{1}{2}}-\frac{L_{1}L_{2}}{r^{2}}.
\end{align}
\end{widetext}
However, as we consider that the collision could also occur at $r < R_{b}$, we need to find the expression of $E_{cm}$ for JMN-1 naked singularity spacetime, which can be written as,

\begin{widetext}
\begin{align}
     \left(\frac{ E_{cm}^2}{2m^{2}}\right)_{JMN-1} &= 1+\frac{1}{(1- M_{0}) \left( \frac{r}{Rb}\right)^{\frac{M_{0}}{1 - M_{0}}}}-\frac{1}{1-M_{0}} \times \nonumber \\ &\left(1-(1- M_{0}) \left( \frac{r}{Rb}\right)^{\frac{M_{0}}{1 - M_{0}}}\left(1+\frac{L_{1}^{2}}{r^{2}}\right)\right)^{\frac{1}{2}}\left(1-(1- M_{0}) \left( \frac{r}{Rb}\right)^{\frac{M_{0}}{1 - M_{0}}}\left(1+\frac{L_{2}^{2}}{r^{2}}\right)\right)^{\frac{1}{2}} -\frac{L_{1}L_{2}}{r^{2}}
     \label{jmn1ecm}
\end{align}
\end{widetext}
\begin{figure*}[ht!]
\centering
\subfigure[$M_{0}$ = 0.50, $R_{b}$ = 4.00.]
{\includegraphics[width=7cm]{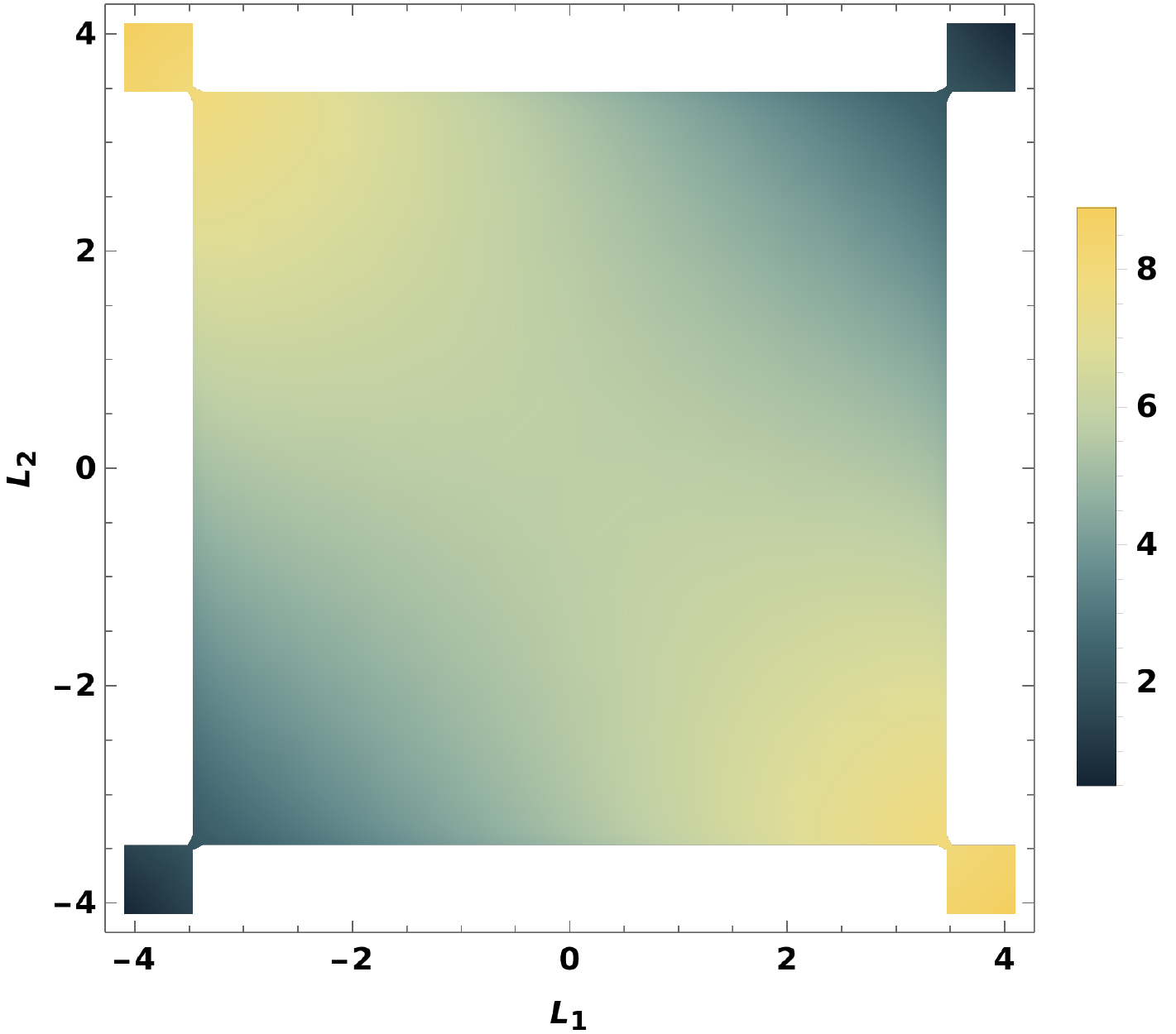}\label{fig:31}}
\hspace{1cm}
\subfigure[$M_{0}$ = 0.55, $R_{b}$ = 3.64.]
{\includegraphics[width=7cm]{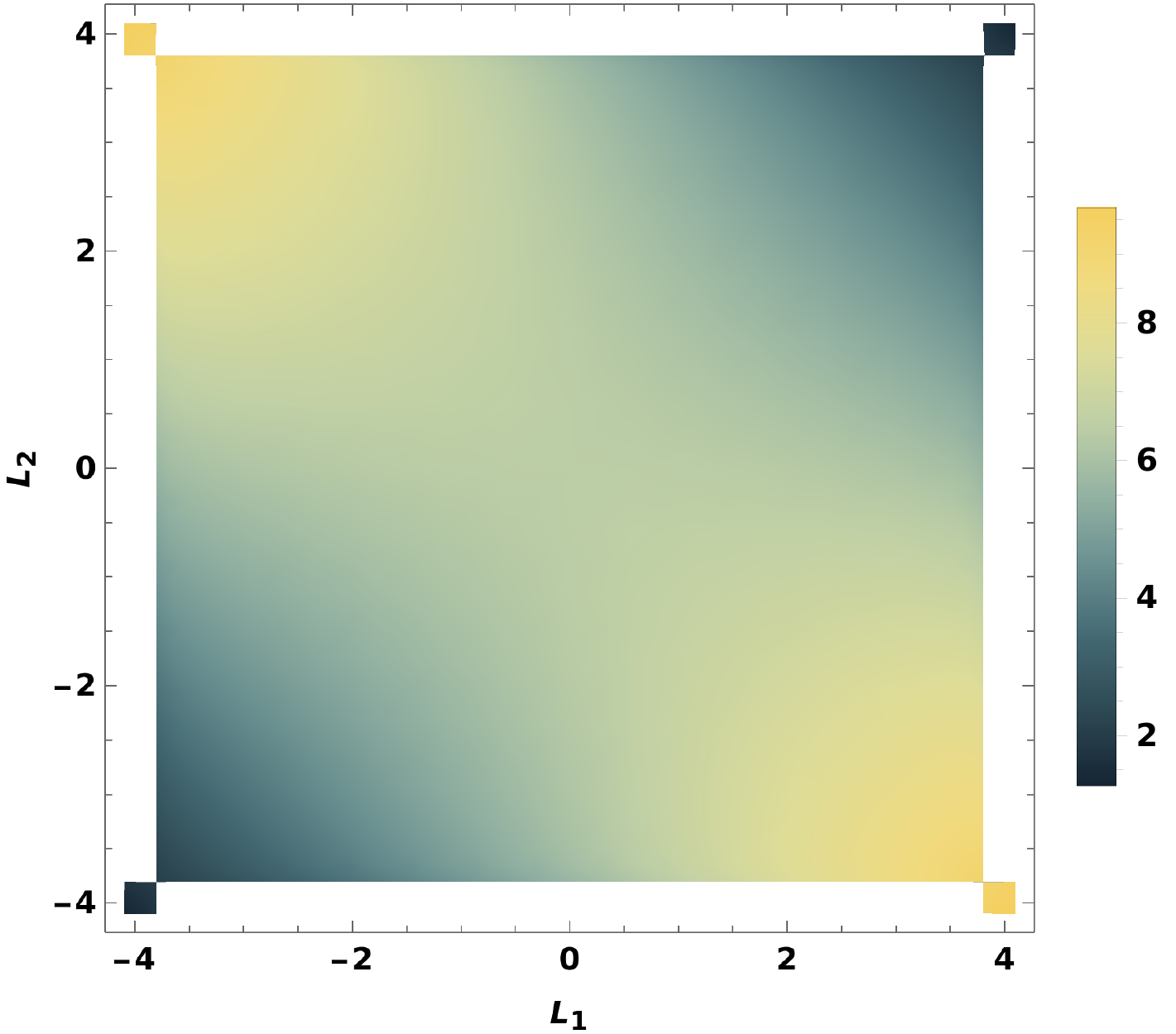}\label{fig:32}}
\subfigure[$M_{0}$ = 0.60, $R_{b}$ = 3.33. ]
{\includegraphics[width=7cm]{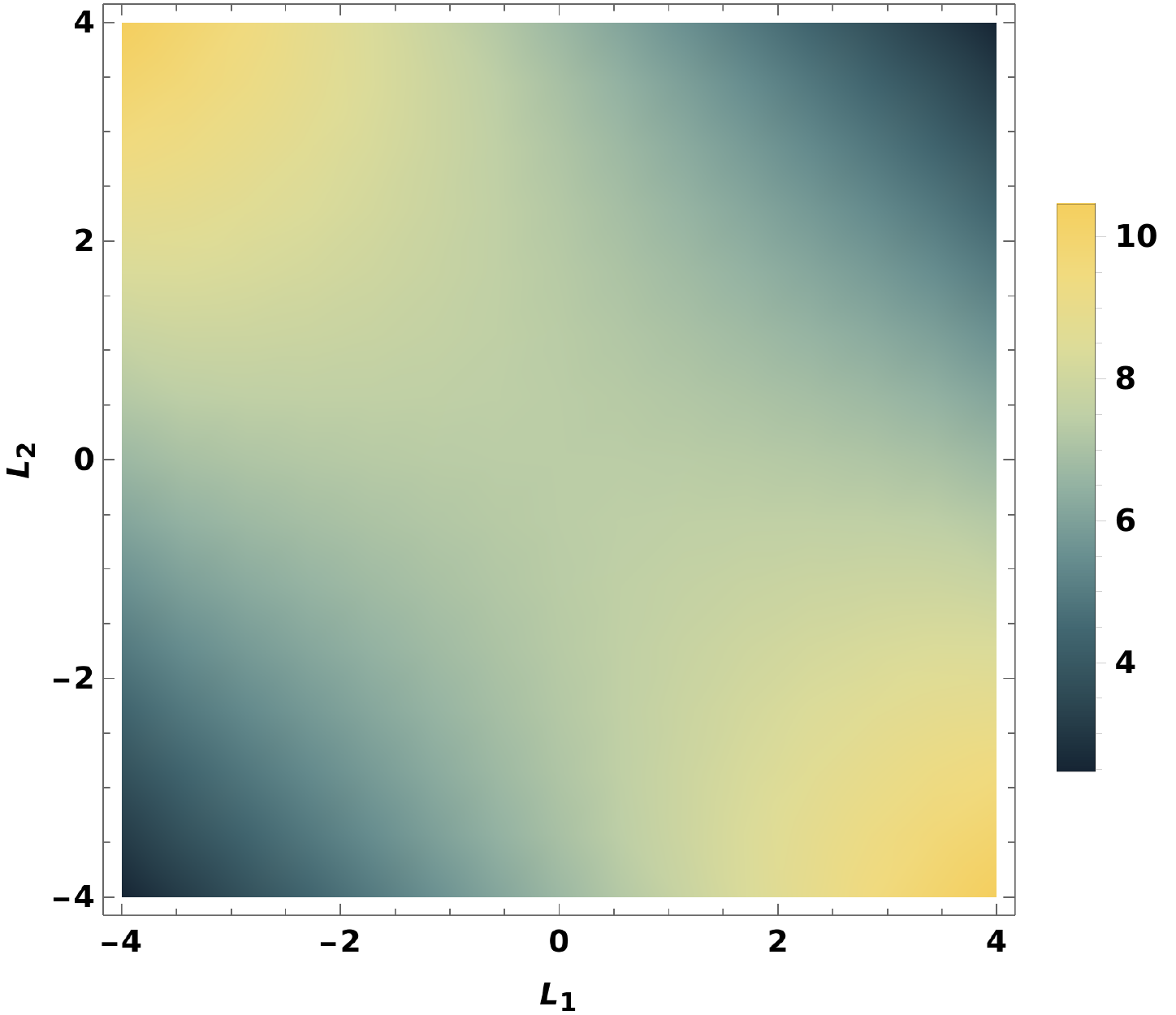}\label{fig:33}}
\hspace{1cm}
\subfigure[$M_{0}$ = 0.65, $R_{b}$ = 3.08. ]
{\includegraphics[width=7cm]{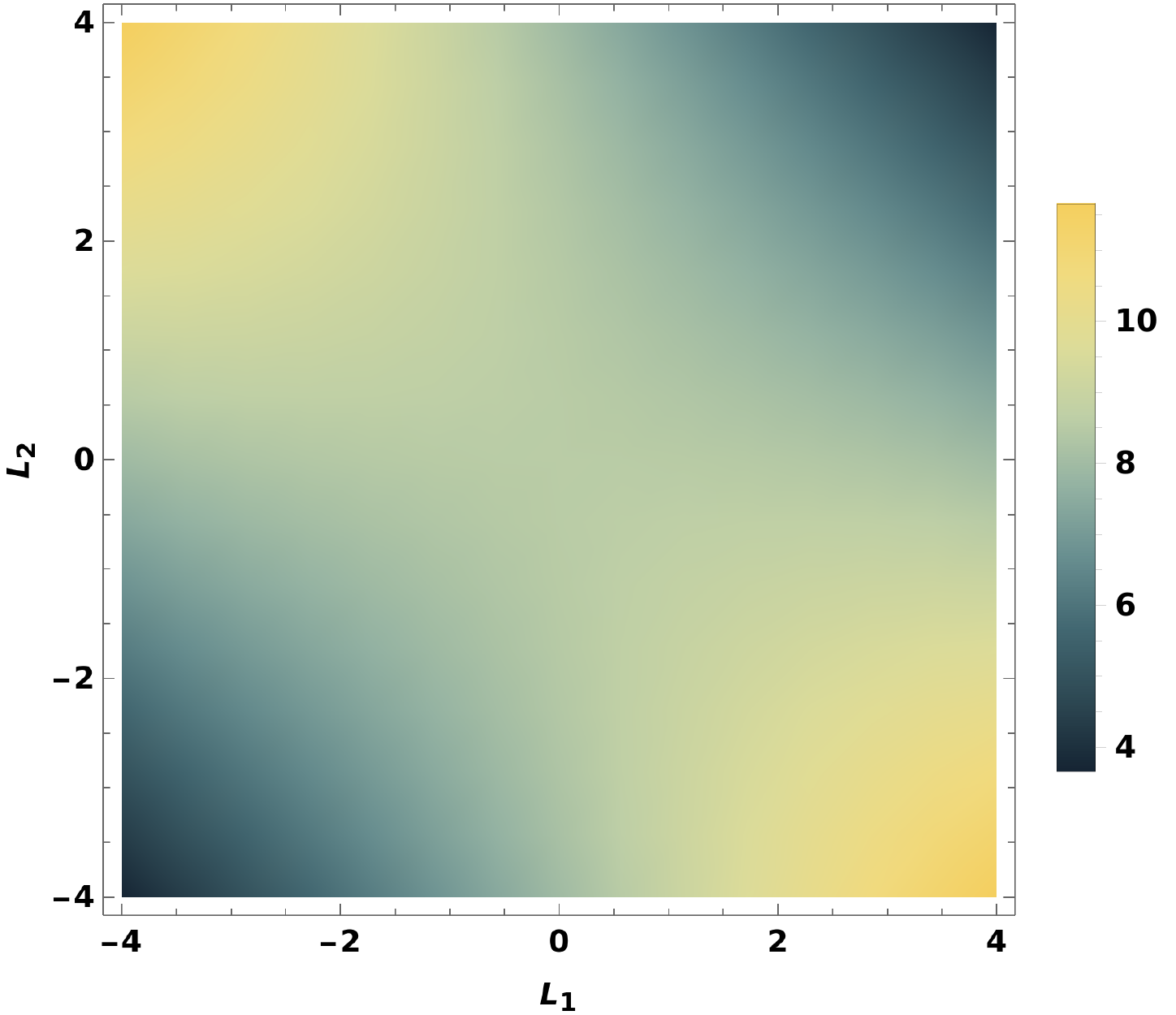}\label{fig:34}}
\hspace{1cm}
 \caption{The above figures show the change in center of mass energy ($E_{cm}$) per unit rest mass with respect to the different angular momentum of two colliding particles. The bar on right side of each figures represent the numerical value of $E_{cm}$. }\label{fig:3}
\end{figure*}
\begin{figure*}[ht!]
\centering
\subfigure[$M_{0}$ = 0.1, $R_{b}$ = 20.]
{\includegraphics[width=7cm]{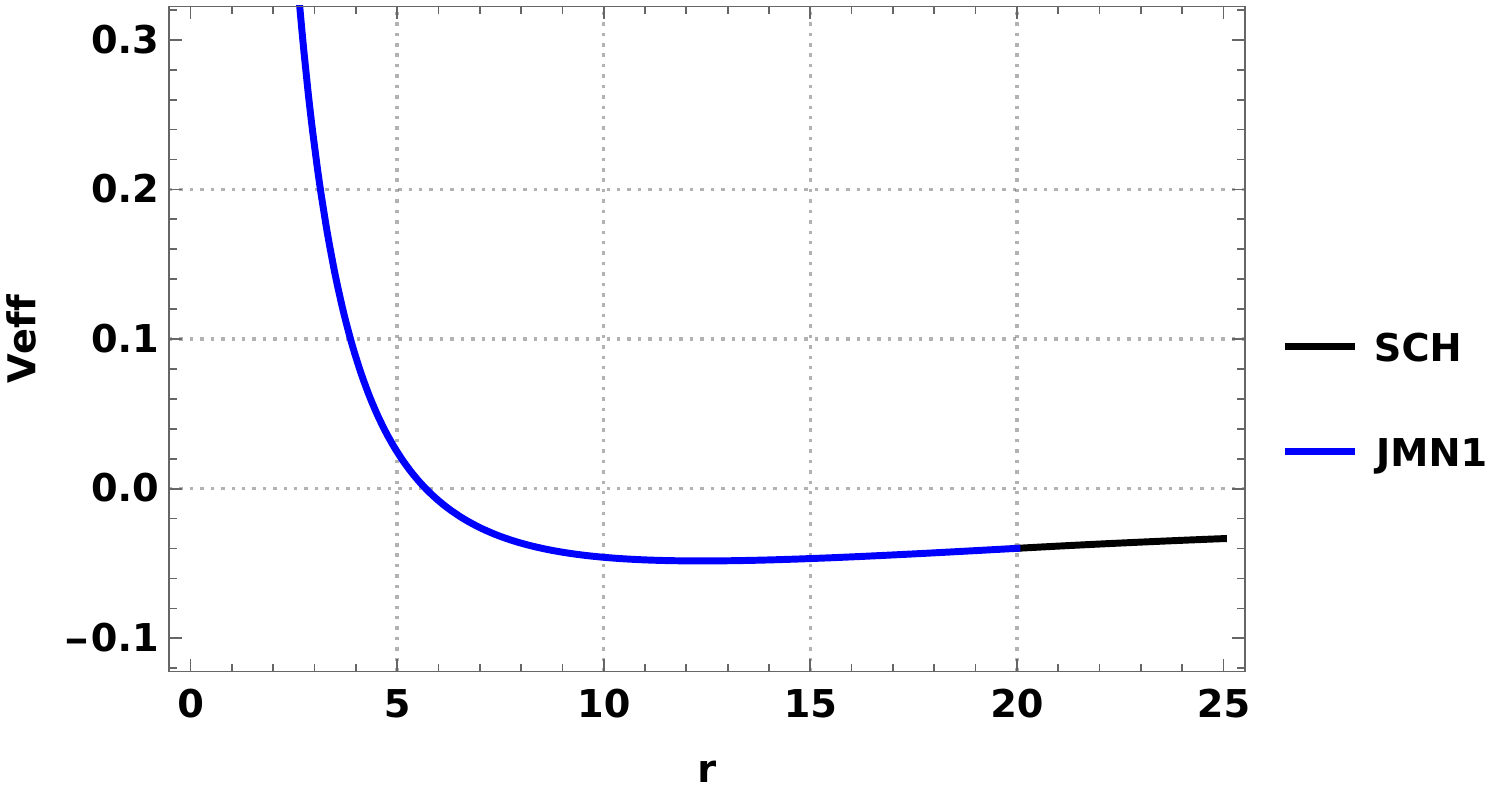}\label{}}
\hspace{1cm}
\subfigure[$M_{0}$ = 0.3, $R_{b}$ = 6.66.]
{\includegraphics[width=7cm]{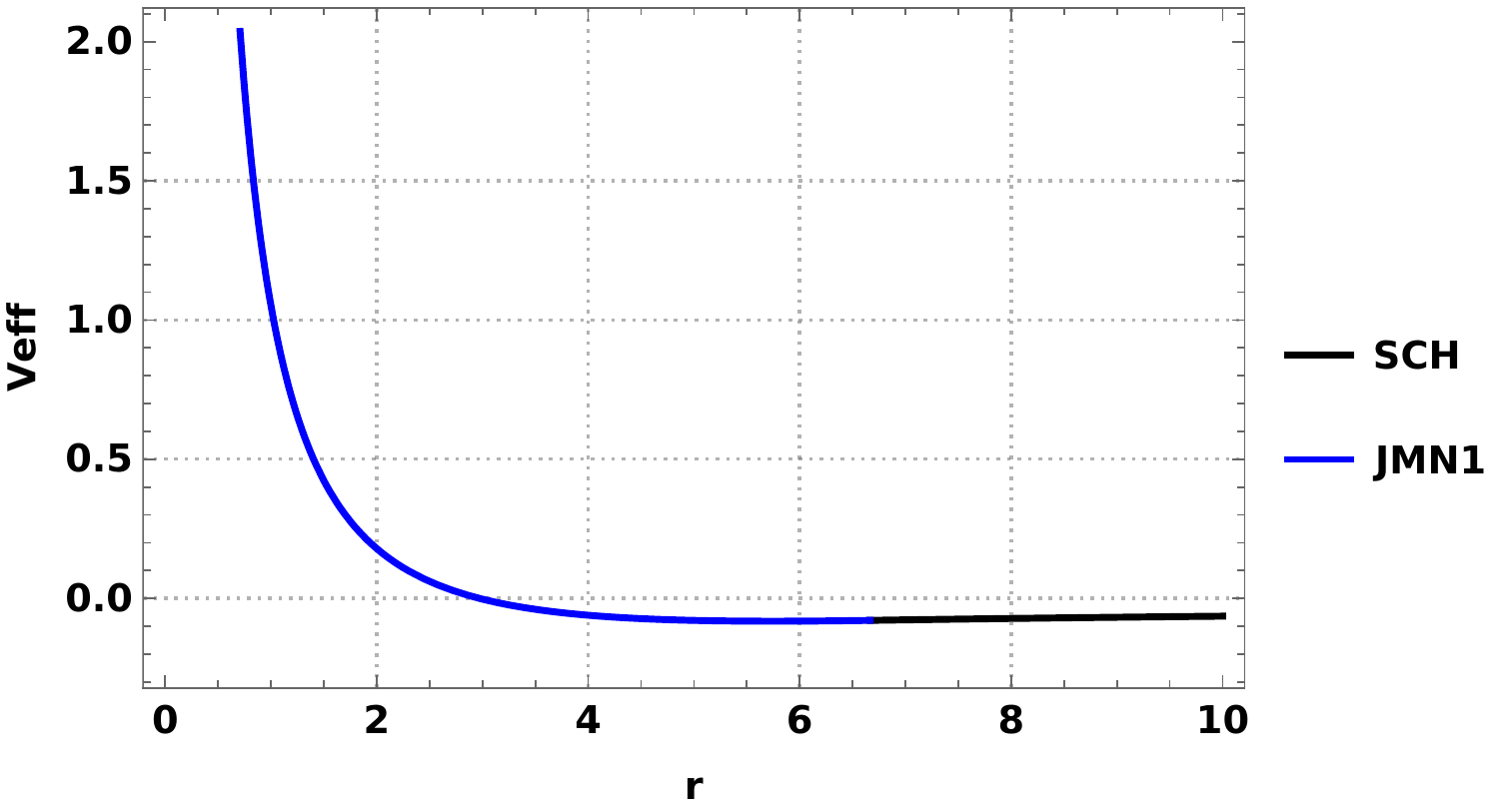}\label{}}
\subfigure[$M_{0}$ = 0.5, $R_{b}$ = 4. ]
{\includegraphics[width=7cm]{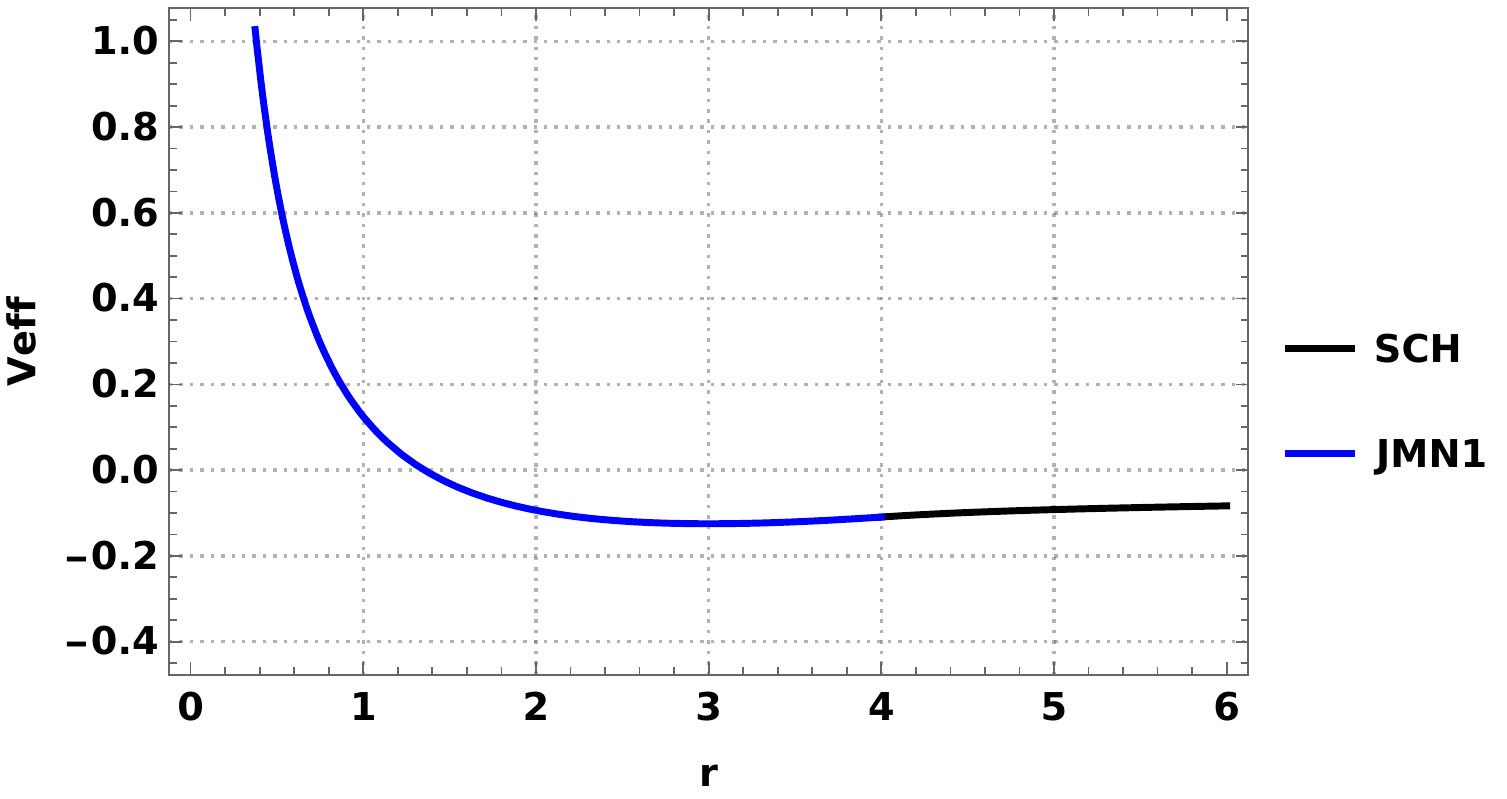}\label{}}
\hspace{1cm}
\subfigure[$M_{0}$ = 0.66, $R_{b}$ = 3.030. ]
{\includegraphics[width=7cm]{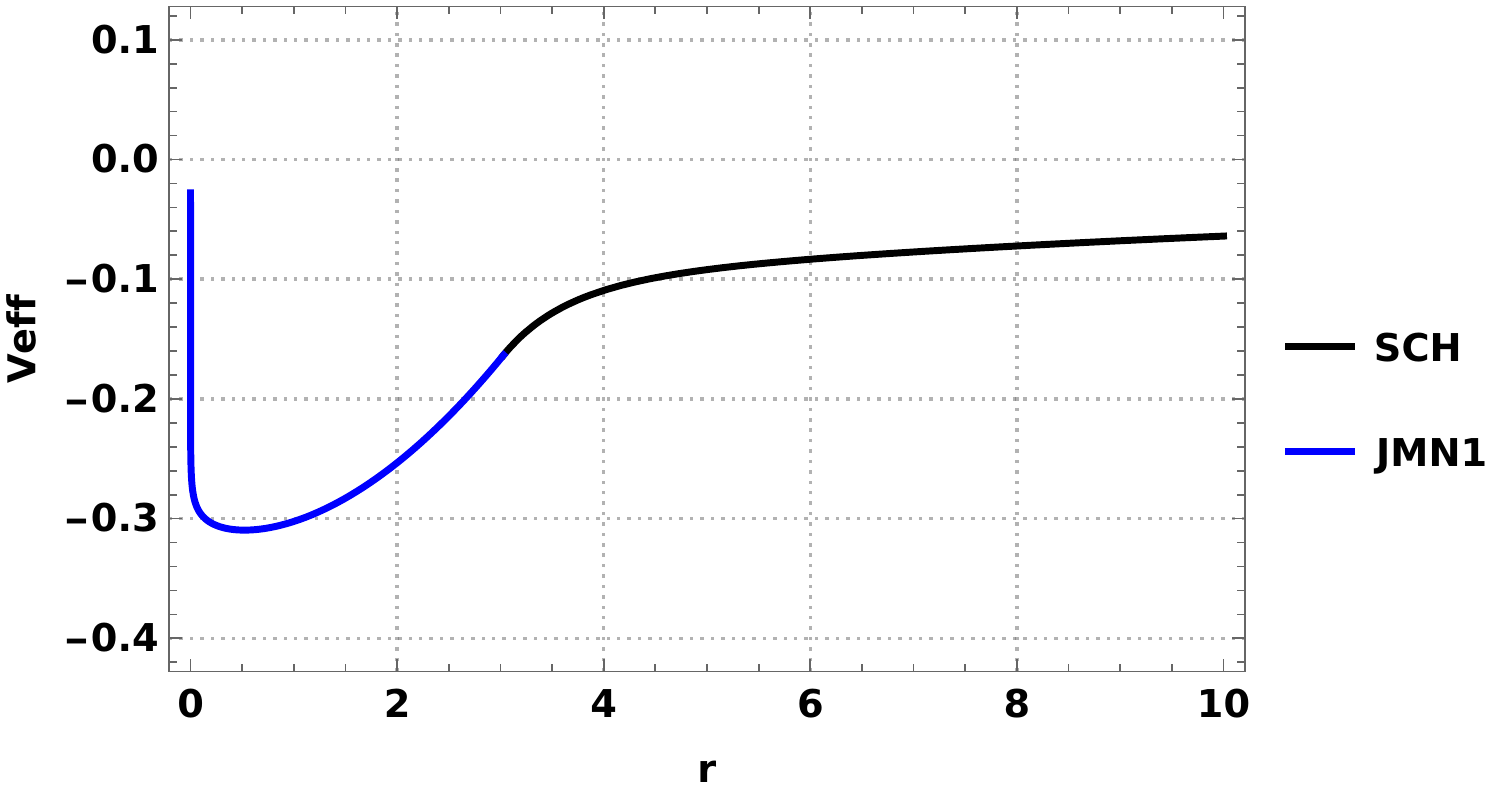}\label{}}
\hspace{1cm}
\subfigure[$M_{0}$ = 0.70, $R_{b}$ = 2.857. ]
{\includegraphics[width=7cm]{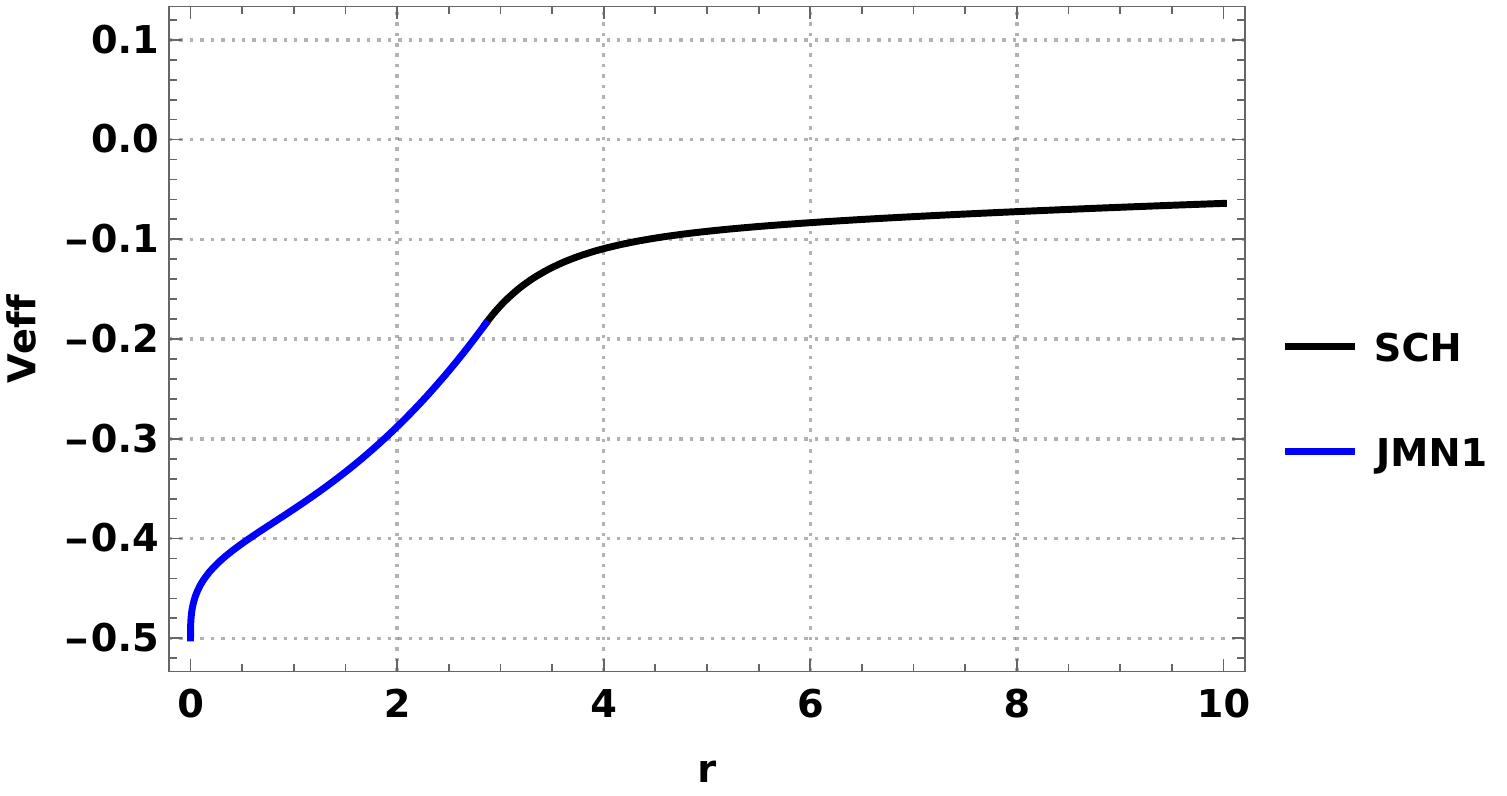}\label{}}
\hspace{1cm}
\subfigure[$M_{0}$ = 0.8, $R_{b}$ = 2.5. ]
{\includegraphics[width=7cm]{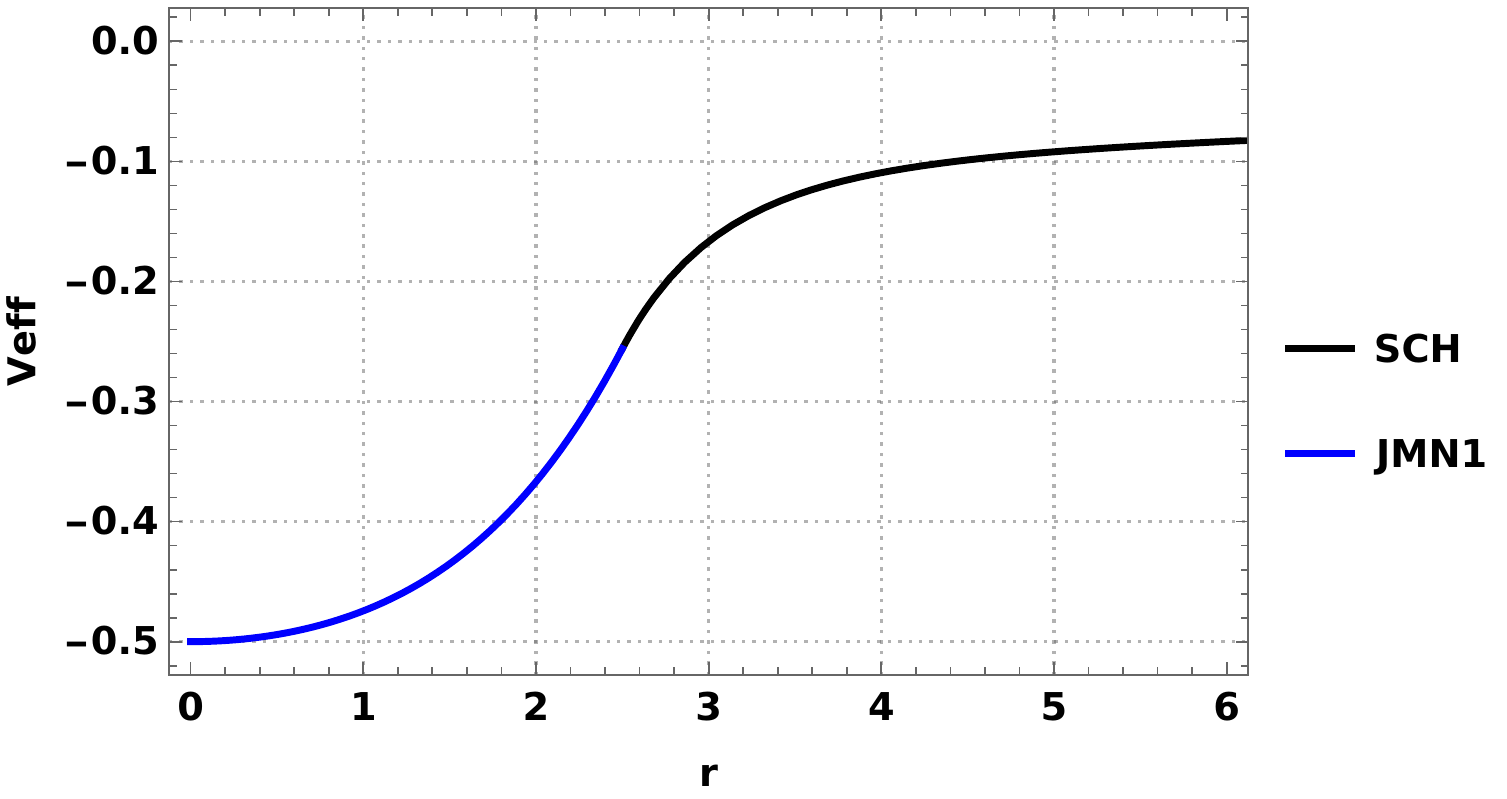}\label{}}
\hspace{1cm}
 \caption{The figure represents an effective potential of internal JMN-1 spacetime (in the blue color) and external Schwarzschild spacetime (in the black color) with different values of $M_0$ and corresponding values of $R_b$. Here, we smoothly matched an interior JMN-1 spacetime with an external Schwarzschild spacetime using the matching conditions at $r=R_b$, \cite{Bambhaniya:2019pbr}. Here, the value of angular momentum is 3.}\label{vefffrodiffm0}
\end{figure*}

From this expression, the energy of both particles after collision in the JMN-1 geometry from the center of mass frame can be evaluated.

At first, to understand the behavior of energy of particle per unit rest mass from center of mass frame of reference $E_{cm}$ with radial distance coordinate, we consider particle collision for $M_{0}=0.55$, which is in the range of allowed values of $M_{0}$ for which particle collision can be considered as the turning point exists at $r<R_{b}$. The results are displayed  in the Fig.\,\ref{fig:radiusandecmjmn1}. The figure shows that as we move away from the singularity, the energy of particle per unit rest mass from the center of mass frame of reference decreases.

However, as we have considered earlier, collision does just take place where the angular momentum of the particles possess values in a certain range.  Thus it is important to see the behavior of $E_{cm}$ with angular momenta $L_{1}$ and $L_{2}$ of two particles. As stated previously, we can consider the collision of particles only when $1/2\,\leq\,M_{0}\,\leq\,2/3$ for the corresponding limiting values of angular momentum. Considering those values, we get plots of energy of two particles per unit rest mass from the center of mass frame with angular momenta of two particles  $L_{1}$ and $L_{2}$ which can be seen in Fig.\,\ref{fig:3}. From this Fig.\,\ref{fig:3}, one can see that the energy from center of mass frame is maximum when both particles possess opposite values of angular momenta. This is because the head-on collision is taking place between the particles as they are approaching each other. This scenario refers to a situation where one particle is moving towards a singularity and another particle is moving away from the same singularity. Then these two particles will collide at the turning point. Also, for the similar values of angular momenta, which suggest the collision of two particles moving in the same direction, $E_{cm}$ is very small.

The center of mass energy $E_{cm}$ will be zero when the magnitude of $L_{1}$ and $L_{2}$ are at their maximum. Notably as the parameter $M_{0}$ increases within the range $\frac{1}{2} \leq M_{0} \leq \frac{2}{3}$), the $E_{cm}$ also increases. We could evaluate the $E_{cm}$ using Eq. (\ref{jmn1ecm}) and Eq. (\ref{angmomjmn}) by considering a neutron with mass, $m = 1.67 \times 10^{-27}$ kg, as the colliding particles and the Sgr A* with mass, $M=4.1 \times 10^{6} M_{\odot}$ as the total mass of the compact object (JMN-1 naked singularity). For the parameter $M_{0}=0.5$, the corresponding value of center of mass energy ($E_{CM}$) of colliding particles can be obtained on the order of $10^{28} eV$, which falls within the Planck energy scale.

\section{Formation of Microscopic Black Holes}
\label{sec_microbh}
According to the theory of general relativity, all forms of energy, including momentum, generate gravity. As a result, it's believed that gravity will become a significant factor at very high energies, specifically when the center of mass energies approach the Planck scales. At Planck energy, the collision of two particles can even create a microscopic black hole \cite{Choptuik:2009ww}.

The four dimensional Planck scale energy is $10^{19} GeV$ or $10^{28} eV$. According to hoop conjecture given by Kip Thorne in 1972, if a large amount of matter and energy, represented by $E$, are compressed into a region, and if a hoop with proper circumference $2\pi R$ can fully encircle this matter in all directions, then in that case a black hole will form if the resulting Schwarzschild radius (calculated as $R_{s} = 2GE/c^{4}$) is larger than the value of $R$. The parameters used in this calculation include Newton's constant ($G$) and the speed of light ($c$) \cite{Choptuik:2009ww}. Since the energy of the collision reaches Planck scales, it can form a microscopic black hole. However, these microscopic black holes are not stable as they decay very rapidly through Hawking radiation \cite{CMS:2010oej}, \cite{Alok:2022xiy}.

One could say that the particles that makeup Hawking radiation are generated by the intense gravitational forces that exist around a black hole. These forces are strong enough to separate virtual particles, which are pairs of particles constantly popping in and out of existence in empty space. When one of these virtual particles is separated from its partner and falls into the black hole, the other particle is free to escape and create the Hawking radiation. Thus through Hawking Radiation, these micro black holes will decay instantaneously \cite{Kovacik:2021qms}.

It is generally believed that miniature black holes decay by releasing elementary particles in the form of a spectrum of energy consistent with that of a black body \cite{Anchordoqui:2002cp}. The energy of these particles can be calculated through Hawking temperature. 
The energy $E$ of Hawking radiation is given by
\begin{eqnarray}
    E = \frac{\hbar c^{3}}{16 \pi GM},
\end{eqnarray}
where $\hbar$ is the reduced Planck constant, $c$ is the speed of light, and $M$ is the mass of the black hole. The particles that will be coming from the decay of microscopic black holes are of much interest and
it would be of interest to consider the energy of such outgoing particles.
The energy of primary particles coming out of microscopic black holes can be obtained using Hawking temperature, as mentioned above. Here the mass of the black hole can be obtained using $M = \frac{E_{CM}}{c^{2}}$.
Since the $E_{CM}$ is equal to the Planck energy, the mass of the microscopic black hole that we obtain will be Planck mass which is in the order of $10^{19} GeV$.
Considering the parameters given above, the energy of these primary particles in the local frame will be in the order of $10^{26} eV$. However, the end results of Hawking emission are not the elementary particles only released by black holes. Others could exist in composite states, while some of them could be unstable hadron that decay  \cite{Arbey:2019mbc}. After all the decay, only the stable particle with a very large half-life will be visible to an asymptotic observer.

It may be worth noting that the energy that we get near JMN-1 singularity is significantly higher than the ultra-high energy cosmic rays that we detect on Earth. However, here we have not considered any other interaction such as Coulomb interaction, Bremsstrahlung radiation, which is electromagnetic radiation emitted when charged particles such as electrons, are accelerated or decelerated by a strong electric field, and any other types of interactions. Further, for an asymptotic observer, the gravitational redshift will also play a significant role in down-scaling the energy. In a physical scenario, these particle interactions and redshift combined will down-scale the energy scale of particles, which might lead us to a possible connection with ultra-high energy cosmic rays. The detailed investigation of the down-scaling of the energy due to various interactions is left for a future work, which may consider the hadronization of Hawking radiation.

\section{Discussions and Conclusions}
\label{sec_conclusion}
We have considered here the phenomena of collision of particles in the vicinity of the JMN-1 naked singularity. We have also derived a numerical value of energy of particles after collision, corresponding to $M_{0} = 0.5$ as a typical case. The conclusions from our investigation can be summarized as below:

\begin{itemize}
   
    \item We considered collision of particles with turning points in this study. For this purpose, we define a range of metric parameter $M_{0}$ for which the particles turn back in the vicinity of the singularity. We find turning of particle is possible within the range  $0 < M_{0} \leq \frac{2}{3}$. For the case of  $ M_{0} >\frac{2}{3}$, the radius at which any particle turns back is imaginary, which suggests that particle plunges inside the singularity. For $M_{0}\,<\frac{1}{2}$, the radius of turning point is outside the boundary radius of JMN-1 metric $R_{b}$. 
    
    \item One can also observe this change from the behavior of the effective potential in fig.\,\ref{vefffrodiffm0}. In this figure, the blue color represents the effective potential in the internal JMN-1 spacetime, while the red color represents the effective potential in the external Schwarzschild spacetime, which is smoothly matched at $r=R_b$.  This figure shows that turning points do not exist for the parameter range $M_0>2/3$. As shown in fig.\,\ref{vefffrodiffm0} (a), (b), (c), and (d), the effective potential diverges arbitrarily for the parameter range $0<M_0\leq2/3$. However, for $4/5>M_0>2/3$, the effective potential has finite upper bounds, as shown in fig.\,\ref{vefffrodiffm0} (e) and (f). 

    \item We find values of radial distance coordinate at which particle turns for different values of $M_{0}$. This radial distance is $r=4.00M$ for $M_{0}=0.50$, $r=3.23M$ for $M_{0}=0.55$, $r=2.43M$ for $M_{0}=0.60$, and $r= 1.31M$ for $M_{0}=0.65$. Here, $r_{min}$ would be the smallest value of radial distance coordinate at which particle can turn back. For $r<r_{min}$, there does not exist turning points as particles plunge inside the singularity. We considered collision of particles for these values of radial distance. 
    
    \item From the range of JMN-1 metric parameter $\frac{1}{2} \leq M_{0} \leq \frac{2}{3}$ and values of radial distance coordinates, we define corresponding values of angular momentum $L$ of two particles approaching the central high density region moving along the geodesic. Finally we studied collision of two particles between the range $\frac{1}{2} \leq M_{0} \leq \frac{2}{3}$ for corresponding values of angular momentum $L$ and radial distance coordinate $r$. In Fig.\,\ref{fig:radiusandecmjmn1} the behavior of centre of mass energy $E_{CM}$ of particles after collision with respect to the radial distance is shown. It can be seen that centre of mass energy decreases as the radial distance increases which is because the gravitational influence of singularity is weak at large distances. 

    \item Another point which should be noted is, in case of black holes, the presence of event horizon leads to requirement of comparatively more fine tuning conditions for the phenomena of particle collision in the vicinity of the high curvature region \cite{Banados:2009pr}. However, in the present case it is seen that without such fine tunings high energy collision of particles occur in the vicinity of the singularity because of absence of horizon.
    
    \item Recently Broderick et al \cite{Broderick:2024vjp} have shown that a wide range of naked singularity models exhibit inner turning points for timelike geodesics for various ranges of values in the corresponding parameter spaces. This leads to the formation of an accretion-powered photosphere within the shadow region of the naked singularity. Consequently, any accretion shock associated with such singularities must appear inside the photon sphere. However, observations of Sgr A* and M87* by the Event Horizon Telescope (EHT) suggest that the accretion flow remains coherent and organized up to the photon sphere. As a result, most types of naked singularity models, except for JMN-1 and JNW naked singularities (classified as type P0j), can be generally ruled out \cite{Broderick:2024vjp}. These exceptions lack the characteristic inner turning points for timelike geodesics before reaching the singularity, making it challenging to detect accretion-driven shocks or photospheres within their shadows using this method.

\item Gravity plays a crucial role at extreme energies, particularly near the Planck scale, where particle collisions can create microscopic black holes \cite{Choptuik:2009ww}. These black holes decay rapidly via Hawking radiation, releasing particles with energies up to $10^{26}$ eV \cite{CMS:2010oej, Alok:2022xiy}. Other potential sources of such extreme energy include blazar jets, gamma-ray bursts, primordial black hole evaporation, and topological defects \cite{Arbey:2019mbc}. However, interactions such as hadronization, Bremsstrahlung radiation, and gravitational redshift can significantly downscale this energy. Notably, energy estimates near the JMN-1 singularity exceed those of high-energy cosmic rays, suggesting a potential connection that requires further investigation.

 
\end{itemize}

We thus find here that massive compact objects can act as natural particle accelerators; not only can they accelerate the particles at an energy that is next to impossible on earth, but also these accelerators can possibly answer some of the most intriguing underlying questions about the ultra-high energy phenomena in the universe. For future work, we will deal with the phenomenology of these particles and how their energies will change with various interactions and processes.

\section{ACKNOWLEDGMENTS}
P. Bambhaniya acknowledge support from the São Paulo State Funding Agency FAPESP (grant 2024/09383-4). V. Patel acknowledges the support of the Council of Scientific and Industrial Research (CSIR, India,
Ref: 09/1294(18267)/2024-EMR-I). The authors express their gratitude towards Prof. P C Vinodkumar, Tapobroto Bhanja, and Saurabh for their valuable suggestions and comments.

\end{document}